# Ethical Statistical Practice and Ethical AI

Rochelle E. Tractenberg
Collaborative for Research on Outcomes and -Metrics, and Georgetown University, Building D, Suite 207, 4000 Reservoir Road NW, Washington, DC 20057

**Abstract**
Artificial Intelligence (AI) is a field that utilizes computing and often, data and statistics, intensively together to solve problems or make predictions. AI has been evolving with literally unbelievable speed over the past few years, and this has led to an increase in social, cultural, industrial, scientific, and governmental concerns about the ethical development and use of AI systems worldwide. The ASA has issued a statement on ethical statistical practice and AI (ASA, 2024), which echoes similar statements from other groups. Here we discuss the support for ethical statistical practice and ethical AI that has been established in long-standing human rights law and ethical practice standards for computing and statistics. There are multiple sources of support for ethical statistical practice and ethical AI deriving from these source documents, which are critical for strengthening the operationalization of the "Statement on Ethical AI for Statistics Practitioners". These resources are explicated for interested readers to utilize to guide their development and use of AI in, and through, their statistical practice.

## 1. Introduction

Artificial Intelligence (AI) is defined as:

"… a field, which combines computer science and robust datasets, to enable problem-solving. It also encompasses sub-fields of machine learning and deep learning... AI algorithms … seek to create expert systems which make predictions or classifications based on input data."[1]

AI very often arises at the intersection of computing and statistics (some AI does not use data explicitly or implicitly, or any statistical methods or modeling). These fields each have longstanding ethical practice standards - the *Code of Ethics and Professional Conduct* of the Association of Computing Machinery (ACM, 2018) and the *Ethical Guidelines for Statistical Practice* of the American Statistical Association (ASA, 2022). These describe the ethical practice of any person at any level of training or job title who utilizes computing (ACM) or statistical practices (ASA). It has been argued that AI can be developed *and deployed* ethically when the ethical practice standards of each of these fields are followed (Tractenberg, 2024).

---

[1] https://www.ibm.com/topics/artificial-intelligence#:~:text=At%20its%20simplest%20form%2C%20artificial,in%20conjunction%20with%20artificial%20intelligence.

In 2024, the ASA Committees on Professional Ethics and Computing jointly drafted a "statement on ethical AI" for statistics practitioners. A key characteristic of this document is that it needed to be brief. The themes determined for organizing this brief statement are Accountability, Transparency, and Fairness. To expand the authority behind the Statement, and to contextualize its recommendations within the wider guidance available, this paper discusses the explicit backing for the new ASA Statement on Ethical AI from the ASA Ethical Guidelines for Statistical Practice, as well as from the ethical practice standards for computing, the Association of Computing Machinery's Code of Ethics and Professional Practice (Association of Computing Machinery, 2018), and global consensus based human-rights oriented guidance from IEEE & the Toronto Declaration (both 2018) for ethical AI.

**1.1 Dimensions of Ethical AI**
Eight dimensions were discovered to be common to "ethical AI" based on 36 statements identifying 47 unique elements that are necessary for "ethical AI" published as of 2020 (Fjeld et al. 2020):

- Privacy
- accountability
- safety and security
- transparency
- fairness and non-discrimination
- human control
- professional responsibility
- promotion of human values

This 2020 paper analyzed 36 different documents, all of which were created to outline what the writers consider to be "ethical AI". It does not seem that any of the authors of these 36 documents had read any of the other authors' work on the topic – including the ASA and ACM ethical practice standards (neither of which was cited in the 2020 analysis). Nevertheless, these 36 documents articulated a total of 47 unique "ethical AI" principles which were grouped according to eight key themes or principles: privacy, *accountability*, safety and security, *transparency* and explainability, *fairness* and non-discrimination, human control of technology, professional responsibility, and promotion of human values. The three themes included in the ASA Statement on Ethical AI are italicized here to contextualize them in this much wider array of guidance on "ethical AI".

For concision, the 2024 ASA Statement on Ethical AI features just Accountability, Transparency, and Fairness. These are undoubtedly critical elements of ethical practice in statistics and data science and computing. Fairness is an important human-rights oriented theme, but is challenging to include in a "statement" because it is challenging to identify to whom exactly AI is supposed to be "fair". Note that the Fjeld et al. analysis refers to "fairness and non-discrimination", which is consistent with Universal Human Rights (Toronto Declaration, 2018). The ASA Statement does not direct the reader to resources that can help them to actually implement the recommendations in the statement (beyond the ASA Ethical Guidelines for Statistical Practice), and due to the concision of the statement, these three themes may mislead ethical statistical practitioners into believing there "is not much to" ethical AI. Concision is not actually helpful in this case.

The ASA Ethical Guidelines for Statistical Practice (2022) define / apply to "statistical practices", including "*activities such as: designing the collection of, summarizing, processing, analyzing, interpreting, or presenting, data; as well as model or algorithm development and deployment*." These 2022 Guidelines are explicitly geared towards those who develop statistical methods and algorithms, *and* those who utilize statistical methods and algorithms. The Guidelines do not specify or suggest limitations on the size of data sets, complexity of models, nor any other feature that the Statement on Ethical AI suggests might lead practitioners to consider beyond or outside the scope of these statistical practices. The new ASA Statement on Ethical AI briefly alludes to ASA Ethical Guidelines, as well as the recognition that there are two distinct dimensions of "ethical AI" for statistical practitioners:

- Ethical *use* of AI
- Ethical *development* of AI.

In fact, the ASA Ethical Guidelines (2022) explicitly apply to every individual and organization engaging in statistical practice irrespective of job title, level, or field of degree. The ACM Code of Ethics (2018) simliarly apply to "anyone using computing in an impactful way". These two ethical practice standards exist to promote ethical decision making by all who engage in "statistical practice" (ASA) and "computing" (ACM). The ASA and ACM ethical practice standards both argue that users of the tools and techniques of these fields – irrespective of training, job title, and career stage – have an obligation to use the tools and techniques of the fields ethically. Any creator, or user, of AI who follows these ethical practice standards (which both assert they should!) will be engaging in ethical AI – creation and use (Tractenberg, 2024). This point should have been made more clearly in the Statement.

## 1.2 Tight support by ASA Ethical Guidelines for Statistical Practice for Ethical AI

The ASA Ethical Guidelines for Statistical Practice (2022) articulates eight (other) Principles and a total of 72 elements including the Appendix with 12 items. Everyone who utilizes statistical practices should be aware of, and following, each of these Principles whenever they are using statistical practices, according to the ASA Ethical Guidelines. The ASA specifically tasks its Committee on Professional Ethics with maintaining and promulgating these Guidelines, yet the ASA as an organization prioritizes concision in the Statement on Ethical AI over encouraging statistics pracitioners who develop or use AI to consider when and where these Guidelines are actually relevant wherever statistical and data science practices are utilized relating to AI.

In fact, an examination of the alignment of elements within the ASA Ethical Guidelines for Statistical Practice (ASA 2022) and the eight common principles for Ethical AI synthesized by Fjeld et al. (2020) highlights the considerable representation within the ASA Ethical Guidelines on all the eight core principles of "ethical AI" - none of which is reflected in the new Statement on Ethical AI. This table also shows that both "ethical AI" and "ethical statistics and data science" involve so much more than just "accountability", transparency", and "fairness". For example, Fjeld et al (2020) identified privacy protection and human values promotion as two specific –different – principles,

Degrees of Freedom Analysis (Tractenberg, 2022) was used to explore the alignment between the ASA Ethical Guidelines for Statistical Practice and the eight Principles for Ethical AI synthesized from 36 documents by Fjeld et a. (2020). In each cell in Table 1 below, the specific elements of the ASA Ethical Guidelines appear where they support the eight Core Principles of Ethical AI (Fjeld et al. 2020).

**Table 1:** Alignment of Principled AI (Fjeld et al. 2020) and ASA Ethical Guidelines (ASA, 2022) 2 august 2024. THREE GREY COLUMN headings = ASA STATEMENT ON ETHICAL AI 'PRINCIPLES'

| 2022 ASA Ethical Guideline Principles (# of elements) | Privacy | Accountability | Safety/ security | Transparency/ explainability | Fairness/ non-discrimination | Human control | Professional responsibility | Promotion of human values |
|---|---|---|---|---|---|---|---|---|
| **A. Professional Integrity & Accountability (12)** | | A; A1; A2; A7; A10; A11; A12 | A2; A3; A4; A7; A8; A11; A12 | A2; A4; A5;A7 | A2; A3; A4; A5; A8; A12 | A1; A4 | ALL A | A2; A3; A4 A8; A12 |
| **B. Integrity of data and methods (7)** | B4 | B1; B2; B3; B5; B6 | B1; B2; B3; B4; B5; B6 | B1; B2; B3; B5; B6 | B2; B3;B7 | ALL B | ALL B | B2; B6 |
| **C. Responsibilities to Stakeholders (8)** | C7 | C1; C2 | C4;C7 | C1; C2; C3; C4; C5 | C2; C3; C6; C7 | C1; C2; C4; C5; C7 | ALL C | C2; C7; C8 |
| **D. Responsibilities to research subjects, data subjects, or those directly affected by statistical practices, Data Subjects, or those directly affected by statistical practices (11)** | D4; D5; D7; D9; D10; D11 | D1; D2; D3; D6; D7; D8; D9; D10; D11 | D1; D2; D3; D4; D5; D7; D9; D10; D11 | D2; D7; D8; D9 | D4; D5; D6; D10; D11 | ALL D | ALL D | D1; D2; D4; D5; D6; D9; D10; D11 |
| **E. Responsibilities to members of multidisciplinary teams (4)** | | E2; E4 | E2 | E3 | | E4 | ALL E | E3; E4 |
| **F. Responsibilities to Fellow Statistical Practitioners and the Profession (5)** | | F3 | | F2 | F3 | | ALL F | |
| **G. Responsibilities of Leaders, Supervisors, and Mentors in Statistical Practice (5)** | | ALL G | G1; G2 | G5 | G5 | | ALL G | G1; G2 |
| **H. Responsibilities regarding potential misconduct (8)** | | H1; H2; H5 | ALL H | H2; H5 | H3; H5 | ALL H | ALL H | H2; H5; H8 |
| **APPENDIX: Responsibilities of organizations/institutions (12)** | | APP 1; 2; 4; 6; 8; 9; 10; 11 | APP 1; 2; 6; 7 | APP 1; 2; 4; 5; 7; 8; 9; 10; 11 | APP 1; 2; 4; 5; 6; 7; 10; 11; 12 | ALL APPENDIX | APP 1; 2; 3; 4; 8; 9; 10; 11 | APP 1; 5; 6; 7; 10; 11 |

Table 1 shows how minimal is the coverage within the Statement on Ethical AI by the three elements chosen for the Statement, particularly as compared to the ASA Ethical Guidelines for Statistical Practice (ASA 2022). Moreover, Table 1 also shows that the focus on "concision" prioritized in the Statement on Ethical AI dramatically underrepresents what a consensus from 36 different documents - even though these were analyzed in 2020 - represents as the core for "ethical AI". Since the Statement on Ethical AI does not relate itself in any way to the Ethical Guidelines for Statistical Practice beyond including a link, the ASA Statement on Ethical AI reflects limited authority to assert its contents as obligations – unlike the Guidelines. If the Statement had relied on the Guidelines, then the ASA would have been a) internally consistent; and b) able to show how and why ethical AI –development *and use* – is an ethical obligation - for all of the statistical practitioners, just like the Ethical Guidelines assert.

## 2. Other important considerations for "ethical AI"

As argued above, a primary deficit in the ASA Statement on Ethical AI is its failure to adequately leverage the ASA Ethical Guidelines for Statistical Practice. Table 1 shows clearly that the three areas included in the AI Statement only cover a very small subset of what is globally considered requisite for "ethical AI" by Fjeld et al. (2020). Moreover, the ASA Ethical Guidelines comprise in "statistical practice", includes "activities such as: designing the collection of, summarizing, processing, analyzing, interpreting, or presenting, data; as well as *model or algorithm development and deployment*" (ASA 2022, emphasis added to highlight the inclusion within these Ethical Guidelines of AI use and development). The failure to leverage the ASA Ethical Guidelines, by articulating that the obligation to develop and deploy AI ethically is also a responsibility of the ethical statistical practitioner, is a considerable limitation on the utility of the ASA Statement on Ethical AI.

### 2.1 Other Ethical Guidance for AI use and development that was excluded from the Statement
Additional recent important resources should also be leveraged to bring any operationalization and/or support for the statistical practitioner who seeks to engage ethically in the use or development of AI. All of the arguments about the ASA Ethical Guidelines and their potential to have strengthened the ASA Statement on Ethical AI also relate to the ACM Code of Ethics. Since there is considerable alignment (all but 2 elements) between the ASA and ACM ethical practice standards (Tractenberg, 2022), if the Statement had incorporated any aspect of, or appealed to, the ASA Ethical Guidelines, then it could have also leveraged the authority of the ACM in making a brief argument that ethical AI comes from following ethical practice standards like that of the ASA.

Like the ASA Ethical Guidelines, "The (ACM) Code is designed to inspire and guide the ethical conduct of *all computing professionals*, including current and aspiring practitioners, instructors, students, influencers, and anyone who uses computing technology in an impactful way." (ACM 2018; emphasis added) As noted above, the ACM Code of Ethics overlaps almost entirely with the ASA Ethical Guidelines (Tractenberg, 2022A, 2022B). The ACM Code has four principles:
- General Ethical Principles (7 items, including "respect human rights")
- Professional Responsibilities (9 items)
- Professional Leadership Principles (7 Principles)
- Compliance with the code (member-facing; 2 items)

The global initiative to develop the IEEE Ethically Aligned Design (EAD) (2018) guidance sought to "establish frameworks to guide and inform dialogue and debate around the non-technical implications of these technologies." (i.e., AI & computing). The stated goal opf EAD was to promote: "The ethical design, development, and implementation of ...intelligent and autonomous technical systems" (p. 6). The EAD initiative identified five General Principles (that ethical design, development, and implementation of intelligent and autonomous technical systems should be guided by):
- Ensure they do not infringe on internationally recognized human rights
- Prioritize metrics of well-being in their design and use
- Ensure that their designers and operators are responsible and accountable
- Ensure they operate in a transparent manner
- Minimize the risks of their misuse

It is worth noting that the IEEE EAD does not add uniquely to the ACM code because both recognize obligations to protect "human rights", which are not explicit in the ASA guidance (but have been suggested for the next quinquennial ASA Ethical Guidelines revision). The IEEE EAD document does not refer to the ACM Code, which had originally been published in 1992 (so was available for EAD developers to utilize). In fact, exploration of the 36 "ethical AI" documents analyzed by Fjeld et al. (2020) demonstrates that they do not refer to each other; none of the ones published prior to 2018 were cited by EAD (2018), nor did any cite the ACM Code (2018).

The Toronto Declaration outlined Universal Human Rights (2018) relating specifically to machine learning, one narrow type of AI. The Toronto Declaration notes that: "States have obligations to *promote, protect and respect* human rights; private sector actors, including companies, have a responsibility to *respect* human rights at all times." (emphasis added) In particular, the Universal Human Rights that pertain to machine learning and are legally recognized worldwide are:

- The right to equality and non-discrimination
- Preventing discrimination
- Protecting the rights of all individuals and groups: promoting diversity and inclusion
- Human rights due diligence:
i. Identify potential discriminatory outcomes
ii. Take effective action to prevent and mitigate discrimination and track responses
iii. Be transparent about efforts to identify, prevent and mitigate against discrimination in machine learning systems.
- Accountability of individuals and organizations (including business entities, governments, and multi-country entities).

The Toronto Declaration does not add unique elements to the conversation, it simply summarizes the established human rights laws already in existence (in 2018), and frames them with respect to AI and machine learning. One interesting element that this Declaration can contribute is an elaboration of "human rights due dilligence" as a concrete replacement for the much more vague term "fairness" in the ASA Statement on Ethical AI. The Toronto Declaration was analyzed in the Fjeld et al. 2020 paper, but is not recognized in the ASA or ACM ethical practice standards; nor is a human rights dimension reflected in the ASA Statement on Ethical AI.

**2.2. What is missing in the ASA Statement on Ethical AI**

To summarize, the context for the new Statement on AI from the ASA can be conceptualized in Figure 1. Figure 1 shows that the eight consensus principles of "ethical AI" synthesized from 36 independent documents (Fjeld et al. 2020) make up only a small portion of the total universe of ethical statistical and computing practices; and the concise ASA Statement on Ethical AI includes only three of those eight principles. Simply by explicitly contextualizing the obligations of the ethical statistical practitioner to follow the ASA Ethical Guidelines when developing or deploying AI (rather than citing the Guidelines but providing no conceptual link), the Statement could have engaged readers in both the actual authority to describe ethical obligations of statistical practitioners to engage in ethical AI practices as users and also as developers, and also reinforced the importance of the ASA's own statement of ethical statistical practices.

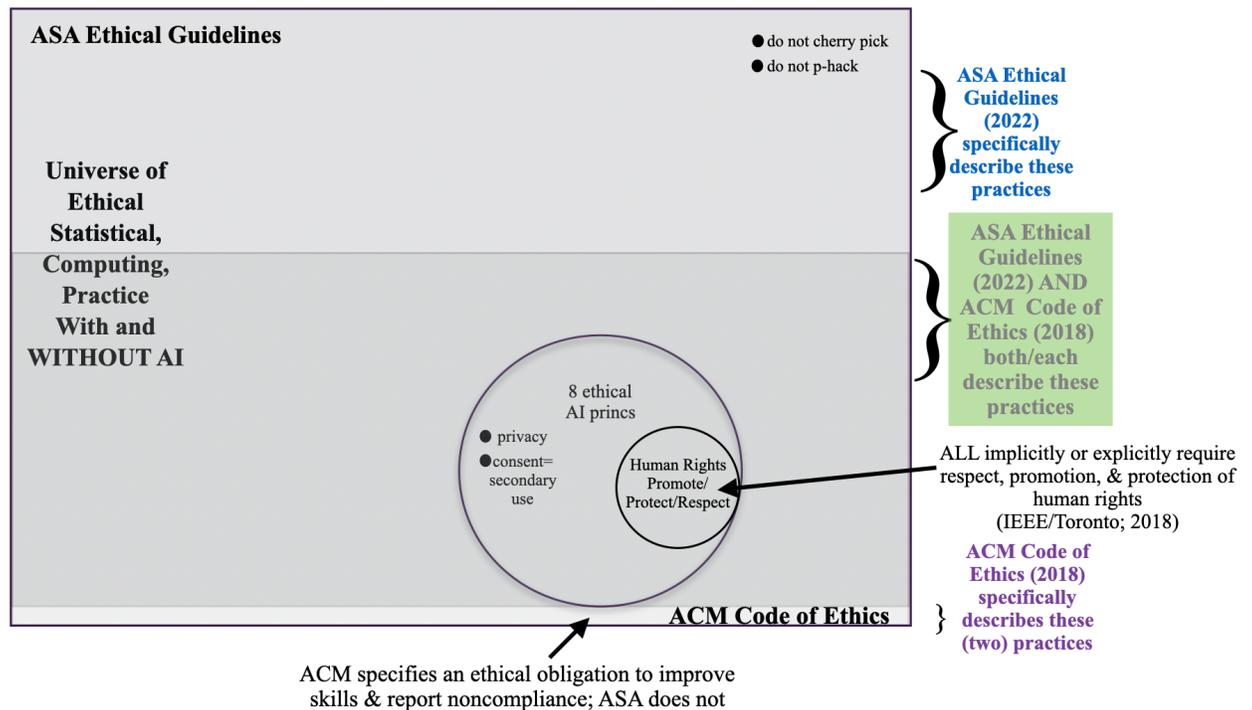

**Figure 1:** Venn diagram of the Universe of Ethical Statistical and Computing Practices with and without AI, highlighting the extensive coverage of the ASA Ethical Guidelines for Statistical Practice and how it covers all of the eight principles of "ethical AI" identified by Fjeld et al. (2020), universal human rights (Toronto Declaration, 2018), and the majority of the ACM Code of Ethics content.

This Figure shows how little of the overall universe of ethical practice in statistics, data science, and computing are captured by the eight principles Fjeld identified as core for "ethical AI". It also shows how far the ASA Statement on Ethical AI could have gone, authority-wise, if it had leveraged the ASA Ethical Guidelines explicitly– because only two elements of the ACM Code are not already addressed with the ASA Ethical Guidelines. Referencing both would have strengthened statistical practitioner awareness of the importance of ethical practice with both statistics and computing.

Moreover, the ASA Ethical Guidelines and ACM Code of Ethics explicitly cover both *contributors to, and users of,* the fields' respective tools and techniques.

## 3. Specific ASA Ethical Guidelines support for use and development of "ethical AI"

While there are clear and compelling arguments that there should have been stronger and more explicit alignment between the Statement on Ethical AI and the ASA Ethical Guidelines, among other key resources for ethical AI, there should also be consideration of what *is* included in the Statement. Clearly, AI has been increasingly utilized in government and public sectors alike, with no likelihood that its use will wane in the forseeable future. As such, it is critical to ensure that the ethical practice standards that exist to guide ethical computing and ethical statistical practices be leveraged to promote the ethical problem-solving, predictions, and classifications referred to in the definition above. The *Toronto Declaration* (2018) was formulated to highlight international law relating to universal and fundamental human rights. Understanding these laws can help to contextualize the importance of ethical AI. The Toronto Declaration describes the obligations of state and private sector actors, respectively. According to human rights laws, "States have obligations to **promote, protect and respect** human rights; private sector actors, including companies, have a responsibility to **respect** human rights at all times." (emphasis added). Thus, AI applications that subvert or undermine these obligations, and those that do not promote, protect, and respect (states) or respect (private sector) fundamental human rights, cannot be characterized as "ethical". This is true whether the AI itself undermines or performs contrary to these obligations and responsibilities, or the uses to which the AI or its outputs are put undermine or perform contrary to these responsibilities.

Statistical practice is defined as "includes activities such as: designing the collection of, summarizing, processing, analyzing, interpreting, or presenting, data; as well as model or algorithm development and deployment." (ASA, 2022). Statistical practitioners may encounter unexpected ethical challenges that arise from engagement with AI that are new, possibly also distinct, from what has been their experience with statistical practice. However, it is also very likely that much of what is defined as statistical practice is -and will continue to be - core for the development and deployment of AI. The term "statistical practitioner" includes all those who engage in statistical practice, regardless of job title, profession, level, or field of degree. Future iterations of the ASA Ethical Guidelines may be even more explicit about the potential for the ethical statistical practitioner to develop and/or deploy AI.

A global consensus is that ethical AI must be *trustworthy*. "Trustworthy" AI requires that it is explainable, equitable, and protective in terms of promoting of privacy, safety, reliability, validity, human-centricity, and accountability. Each of these features of "trustworthy AI" can be promoted by following the ASA Ethical Guidelines for Statistical Practice, as shown in Table 1. The arguments for how the Guidelines support the three elements of ethical AI laid out in the ASA Statement (Accountability, Transparency, and Fairness) are outlined in the next sections.

### 3.1 ASA Ethical Guidelines for Statistical Practice support for *Accountable AI*
For statistical practice that is used in/for AI, the Principles relating to accountability in the practitioner (Principle A) and the data/methods (Principle B) are most directly relevant for "trustworthy" AI. It is important for practitioners to be aware of evolutions in ethical, regulatory, and legal guidance on the development and use of AI systems (Ethical Guideline Principle D.1). Ethical Guideline Principle B, *Integrity of Data and Methods*, articulates obligations to ensure data fitness for purpose and quality, algorithm design, model evaluation, and deployment. For example,

- Ethical Guideline Principle A, *Professional Integrity and Accountability*, outlines the ethical obligations to take responsibility for one's work. Statistical practitioners should apply and maintain their professional competence and keep up-to-date with the latest developments in AI and statistical practice so that they can take responsibility for the

- outputs of AI they develop and deploy. Understanding how AI generates output is important for evaluating decisions based on AI and for supporting accountability.
- Ethical Guideline Principle B, *Integrity of Data and Methods*, describes ways statistical practitioners can (should) promote accountability and integrity in their work with AI.
- Statistical practitioners, particularly those in leadership roles, have a responsibility to ensure an appropriate ecosystem exists that can ensure ongoing and competent assessment, monitoring, and mitigation of detected risks of AI use (Ethical Guideline Principle G; Appendix)

Statistical practitioners and their organizations should audit and justify the AI systems they use or help develop, and be able to document their accountability in terms of the practitioners (Principle A) and the data/methods used (Principle B).

### 3.2 ASA Ethical Guidelines for Statistical Practice support for *Transparent AI*
Ethical Guideline Principles C, *Responsibilities to Stakeholders*, and D, *Responsibilities to Research Subjects, Data Subjects, and Those Directly Affected by Statistical Practices*, outline the importance of transparency in all aspects of practice for stakeholders and those contributing data. As outlined in Principle C, ethical statistical practice involves stakeholder-accessible, coherent communication. Thus, statistical practitioners developing or deploying AI should strive to be able to explain the reasoning and assumptions underpinning any decision rendered by an AI system. In traditional statistical practice it is perhams clearer that the statistics practitioner makes a set of decisions and assumptions that lead to the outputs of modeling or analyses; the same transparency is critical for trustworthy AI. Ethical Guideline Principles C and D are particularly supportive of transparency in AI. Like more tradidional statistical practices, AI systems should generate relevant and meaningful information for stakeholders (Principle C) and other affected parties (Principle D), and this information should include biases, limitations, and potential risks associated with decisions the AI makes or decisions based on the AI system outputs. For example,
- Statistical practitioners have a responsibility to be knowledgeable about biases as well as the limitations of data *and* statistical work based on this data (Ethical Guideline Principle B).
- Fitness of data and AI-driven decisions for purpose, limitations, and biases in the AI system inputs and outputs are as essential as they are with the models utilizing data, and the outputs of such models in traditional statistical practice.
- Effective communication with stakeholders (Principle C), other team members (Principle E, *Responsibilities to Members of Multidisciplinary Teams*), other statistics practitioners (Principle F, *Responsibilities to Fellow Statistical Practitioners and the Profession*), and those directly affected by statistical practice (Principle D) are all essential to transparent and trustworthy AI.

Statistical practitioners (Principle A) and their organizations (Appendix, *Responsibilities of Organizations/Institutions*; Principle G, *Responsibilities of Leaders, Supervisors, and Mentors in Statistical Practice*) should promote transparency to build trust - as well as accountability - into the AI systems they use or develop.

### 3.2 ASA Ethical Guidelines for Statistical Practice support for *Fair AI*
The Toronto Declaration's outline of fundamental and universal human rights (laws) highlights the importance of developing and deploying AI that does not discriminate, exclude, or do other harm, including creating or perpetuating inequity and other social injustices. Ethical statistical practice will yield accurate and unbiased information for stakeholders, but fairness and equity are harder to ensure because any given society is segmented into a diverse collection of subgroups, and fairness to all is sometimes impossible. Nevertheless, it is possible to ensure that statistical and AI practice does not discriminate, exclude, or do other harm while we strive to support stakeholders' informed judgments and decisions. The responsibilities in Principles A and B extend to the assessment and mitigation of risks of biased outcomes, while Principles C and D mitigate misleading stakeholders and disproportionate impacts. For example,

- Ethical Guideline Principle A, *Integrity and Accountability*, states that the ethical statistical practitioner does not knowingly conduct statistical practices that exploit vulnerable populations or create or perpetuate unfair outcomes (A.3).
- Ethical Guideline Principle C outlines the obligations of statistical practitioners to balance individual and collective interests as well as those of diverse stakeholder groups. These responsibilities extend to minimizing risks of exploitation, bias, and other harms that AI can facilitate or create.
- ASA Ethical Guideline Principle D expounds on these obligations thoroughly. For example, the ethical statistical practitioner does not conduct statistical practice that could reasonably be interpreted by subjects as sanctioning a violation of their rights (Ethical Guidelines Principle D element 11). This responsibiltiy extends to the development or use of AI.

## 4. Conclusions

Statistical practices, as outlined in the ASA Ethical Guidelines for Statistical Practice, include core elements of AI and as such, AI should be developed and deployed in accordance with the ASA Ethical Guidelines with due consideration of universal Human Rights and the ACM Code of Ethics and Professional Practice. All those who utilize statistical practices, irrespective of career stage, training, job title, and context should contribute to the development and use of AI in ways that do not discriminate, exclude, or do other harm. The ethical statistical practitioner who engages in the development or use of AI should do more than focus on accountability, transparency, and fairness *of the AI*; these characteristics as well as all the others outlined in the Ethical Guidelines should be followed for communication about the AI and all statistical practice. Without the background, context, and authority explicated in this paper, the Statement alone is unlikely to guide the development and use of AI in, and through, ethical statistical practice. The ASA Ethical Guidelines for Statistical Practice offer many specific elements that can promote ethical development and use of AI, beyond the three elements outlined in the 2024 ASA Statement on Ethical AI, Accountability, Transparency, and Fairness.

## Acknowledgements

The author acknowledges having the opportunity to review the Statement before it was submitted to the ASA Board for approval. The opinions expressed here are the author's own.## References

Association for Computing Machinery (ACM). (2018) *Code of Ethics and Professional Practice.* Downloaded from https://www.acm.org/about-acm/code-of-ethics on 12 October 2018.

American Statistical Association (ASA). (2022). *ASA Ethical Guidelines for Statistical Practice*-revised, downloaded from https://www.amstat.org/ASA/Your-Career/Ethical-Guidelines-for-Statistical-Practice.aspx on 30 April 2018/2 February 2022.

American Statistical Association. (2024, 8 October). American Statistical Association Statement on Ethical AI Principles for Statistical Practitioners. Downloaded from https://www.amstat.org/docs/default-source/amstat-documents/asa-statement-on-ethical-ai-principles-for-statistical-practitioners.pdf on 10 October 2024.

Fjeld J, Achten N, Hilligoss H, Nagy A, Srikumar M. (2020). "Principled Artificial Intelligence: Mapping Consensus in Ethical and Rights-based Approaches to Principles for AI." Berkman Klein Center for Internet & Society, Publication 2020-1. Downloaded from http://dx.doi.org/10.2139/ssrn.3518482 on 25 June 2024.